\documentclass[proceedings]{stacs}
\stacsheading{2010}{191-202}{Nancy, France}
\firstpageno{191}

\usepackage{amsfonts}
\usepackage{moreverb}
\usepackage{graphicx}

\usepackage[dvips]{epsfig}
\usepackage{amsmath}

\def\inline#1:{\par\vskip 7pt\noindent{\bf #1:}\hskip 10pt}
\def\Proof{\par\noindent{\bf Proof:~}}
\def\blackslug{\hbox{\hskip 1pt \vrule width 4pt height 8pt
    depth 1.5pt \hskip 1pt}}
\def\QED{\quad\blackslug\lower 8.5pt\null\par}
\def\inQED{\quad\quad\blackslug}
\def\prob{\Pi}

\begin{document}
\sloppy
\title{Robust Fault Tolerant uncapacitated facility location}

\author[lab1]{S. Chechik}{Shiri Chechik}
\address[lab1]{Department of Computer Science and Applied Mathematics
  \newline The Weizmann Institute of Science
  \newline Rehovot 76100, Israel}
\email{{shiri.chechik,david.peleg}@weizmann.ac.il}
\author[lab1]{D. Peleg}{David Peleg}


\begin{abstract}
In the {\em uncapacitated facility location} problem, given a graph,
a set of demands and opening costs, it is required to find
a set of facilities $R$, so as to minimize the sum of the cost of opening
the facilities in $R$ and the cost of assigning all node demands
to open facilities.
This paper concerns the {\em robust fault-tolerant} version
of the uncapacitated facility location problem (RFTFL).
In this problem, one or more facilities
might fail, and each demand should be supplied by the closest open facility
that did not fail. It is required to find a set of facilities $R$, so as to
minimize the sum of the cost of opening the facilities in $R$ and the cost of
assigning all node demands to open facilities that did not fail,
after the failure of up to $\alpha$ facilities.
We present a polynomial time algorithm that yields a 6.5-approximation
for this problem with at most one failure and a $1.5 + 7.5\alpha$-approximation
for the problem with at most $\alpha > 1$ failures.
We also show that the $RFTFL$ problem is NP-hard even on trees,
and even in the case of a single failure.
\end{abstract}

\maketitle

\section*{Introduction}

\subsection*{The robust fault-tolerant facility location problem}
For a given optimization problem, the robust fault-tolerant version
of the problem
calls for finding a solution that is still valid even when some components
of the system fail. We consider the robust fault-tolerant version of the
{\em uncapacitated facility location (UFL)} problem.
In this problem, given a graph $G$, a demand $\omega(v)$ for every node $v$
and a cost $f(v)$ for opening a facility at $v$,
it is required to find a set of facilities $R$,
so as to minimize the sum of the costs of opening the facilities in $R$
and of shipping the demands of each node from the nearest open facility
(at a cost proportional to the distance).
In the robust fault-tolerant version of this problem (RFTFL),
one or more facilities
might fail. Subsequently, each demand should be supplied by the closest open
facility that did not fail. It is required to select a set of facilities $R$,
so as to minimize the sum of the costs of opening the facilities in $R$
and the costs of assigning all node demands to open facilities
that did not fail, after the failure of up to $\alpha$ facilities.
We present a polynomial time algorithm that yields a 6.5-approximation
for this problem with at most one failure and a $1.5 + 7.5\alpha$-approximation
for the problem with at most $\alpha > 1$ failures.
We also show that the $RFTFL$ problem is NP-hard even on trees,
and even in the case of a single failure.

\subsection*{Related Work}
Many papers deal with approximating the $UFL$ problem, cf.
\cite{cg,acdb,gk,jms,myz,bs}.
The best approximation ratio known for this problem is 3/2, shown by Byrka in
\cite{jb}.

A fault-tolerant version of the facility location problem was first introduced
by Jain and Vazirani \cite{jv}, who gave it an approximation algorithm
with ratio dependent on the problem parameters.
The approximation ratio was later improved by Guha et al.
to 2.41 \cite{gmm} and then by Swamy and Shmoys to 2.076 \cite{ss}.
However, the variant of the problem studied in these papers is different
from the one studied here.
In that version, every node $j$ is assigned in advance to a number
of open facilities, and pays in advance for all of them. More explicitly,
every node $j$ is assigned to $r_j$ open facilities, and its shipping cost
is some weighted linear combination of the costs of shipping its demand
from all the facilities to which it is assigned.
It is required to find a set of facilities $R$ that minimizes the sum
of the costs of opening the facilities in $R$ and the sum of costs of shipping
the demand of each node $j$ from its $r_j$ facilities in $R$.
This approach is used to capture the expected cost of supplying all
clients demand when some of the facilities fail.
In contrast, in our definition a node $j$ does not pay in advance
for shipping its demand from a number of open facilities.
Rather, it pays only for the cost of shipping its demand from the
surviving facility that actually supplied its demand.
Hence our definition for the fault-tolerant facility location problem
requires searching for a set of facilities $R$ that minimizes the sum
of the costs of opening the facilities in $R$ and the costs of assigning
the demands of each node to one open facility that did not fail,
for any failure of up to $\alpha$ facilities.
Our approach is used to capture the worst case cost of supplying all
clients demand when some of the facilities fail.
We argue that our definition may be more natural in some cases,
where after the failure of some facilities, each demand should still
be supplied by a single supplier, preferably the closest surviving
open facility, and each client should pay only for the cost of shipping
its demand from that surviving facility, and not for all the other
(possibly failed) facilities to which it was assigned originally.
On the technical level, the approach taken in \cite{gmm,jv,ss}
is based on applying randomized rounding techniques
and primal-dual methods to the corresponding integer linear program.
This approach does not readily apply to our version of the problem,
and we use a direct combinatorial algorithmic approach instead.

Two other closely related types of problems are the 2-stage stochastic and 
robust optimization problems (cf. \cite{D+05,F+07}).
Both of these models involve two decision stages. In the first stage, 
some facilities may be purchased. This stage is followed by some scenario
depending on the specifics of the problem at hand
(in a facility location problem for example, the scenario may specify 
the clients and their corresponding demands). 
Subsequently, a second stage is entered, 
in which it is allowed to purchase additional facilities 
(whose cost might be much higher than in the first stage). 
In stochastic optimization there is a distribution over all possible scenarios 
and the goal is to minimize the expected total cost. 
In robust optimization the goal is to minimize the cost of the first stage 
plus the cost of the worst case scenario in the second stage.
In contrast with these two models, in our variant the facilities 
must be selected and opened in advance, and these advance decisions
must be adequate under all possible future scenarios.

Billionnet and Costa \cite{bc} showed a polynomial time algorithm for solving
the ordinary (non-fault-tolerant) UFL problem on trees.
In contrast, we show that the fault-tolerant variant RFTFL is NP-hard on trees.

\section{Preliminaries}

Let us start with common notation to be used later on.
Consider an optimization problem $\prob$ over a universe $V$,
which given an instance $I$, requires finding a solution consisting of
a set of elements $R\subseteq V$.
Denote by $C_{\prob}(I,R)$ the cost of the solution $R$ on the instance $I$
of $\prob$. Let $R_{\prob}^*(I)$ denote the optimal solution
to the problem $\prob$ on instance $I$, and let $C_{\prob}^*(I) =
C_{\prob}(I, R_{\prob}^*(I))$ be the cost of the optimal solution.
We denote our algorithm for each problem $\prob$ studied
later by $A_{\prob}(I)$. The solution returned by the algorithm
is referred to as $R_{\prob}^{alg}(I)$ and its cost is
$C_{\prob}^{alg}(I) = C_{\prob}(I, R_{\prob}^{alg}(I))$.

Let us now define the {\em uncapacitated facility location (UFL)}
problem. Let $I=\langle G ,l ,f, \omega\rangle$ be an instance of
the problem, where $G=(V,E)$ is a graph with vertex set
$V=\{1,...,n\}$ and edge set $E$. Each node $v \in V$ hosts a
client in need of service, and may host a facility, providing
service to clients in nearby nodes. Each edge $e \in E$ has a positive
length $l(e)$. The distance $d(u,v)$ between two points
$u$ and $v$ on $G$ is defined to be the length of the shortest
path between them, where the length of a path is the sum of the
lengths of its edges. For each node $v$, let $f(v)$ denote the
{\em opening cost} associated with placing a facility at $v$, and
let $\omega(v)$ denote the demand of the node $v$. The {\em
shipping cost} of assigning the demand $\omega(u)$ of a client $u$
to an open facility $v$ is the product $SC_{u,v} = \omega(u)
d(u,v)$. The shipping cost $SC_{u,R}$ from a set of open
facilities $R$ to a node $u$ is the minimum cost of assigning $u$
to a server in $R$, namely, $SC_{u,R} = \min\{SC_{u,v} \mid v \in
R\}$. Defining the distance $d(v,R)$ between a set of points $R$
and a point $v$ on $G$ to be the minimum distance
between $v$ and any node in $R$, i.e., $d(v,R) = \min_{r \in
R}{d(v,r)}$, we also have $SC_{v,R} = \omega(v) d(v,R)$.

It is required to find a subset $R \subseteq V$ that minimizes the
sum of costs of opening the facilities in $R$ and the shipping
costs from $R$ to all other nodes. This problem can be formulated
as searching for a subset $R \subseteq \{1,...,n\}$ that minimizes
the cost function

\begin{equation}
\label{eq:1}
C_{UFL}(I,R) ~=~ C_{facil}(I,R)+C_{ship}(I,R),
\end{equation}
where
$$C_{facil}(I,R) ~=~ \sum_{r \in R}{f(r)} ~~~~~ \mbox{ and } ~~~~~
C_{ship}(I,R) ~=~ \sum_{u=1}^n{SC_{u,R}}  ~=~ \sum_{u=1}^n
\omega(u)\cdot d(u,R).$$
Given a set $R$ of open facilities and a
facility $r \in R$, let $\varphi(I,r,R)$ denote the set of clients
that are served by $r$ under $R$, i.e., $\varphi(I,r,R) = \{u \mid
d(v,r) \leq d(v,r')~for~every~r'\in R \}$, or in other words, the
nodes $u$ that satisfy $d(u,R) = d(u,r)$, where ties are broken arbitrarily,
i.e., if there is more than one open facility $r$ such that $d(u,R) = d(u,r)$,
then just choose one open facility $r$ that satisfies $d(u,R) = d(u,r)$
and add $u$ to $\varphi(I,r,R)$. (When the set $R$ is
clear from the context we omit it and write simply $\varphi(I,r)$,
or even $\varphi(r)$ when the instance $I$ is clear as well.)

The {\em robust fault-tolerant facility location (RFTFL)} problem is defined as
follows. Each client is supplied by the nearest open facility, and
in case this facility fails - it is supplied by the next nearest
open facility. We would like to find a solution that is tolerant
against a failure of one node. This problem can be formulated as
searching for a subset $R \subseteq \{1,...,n\}$ that minimizes
the cost function

\begin{equation}
\label{eq:2}
C_{RFTFL}(I,R) ~=~ C_{facil}(I,R)+C_{ship}(I,R)+C_{backup}(I,R),
\end{equation}
where $C_{facil}(I,R)$ and $C_{ship}(I,R)$ are defined as above and
\begin{equation}
\label{eq:3}
C_{backup}(I,R) ~=~ \max_{r\in R}\left\{\sum_{v\in\varphi(I,r,R)}
\omega(v)\cdot(d(v,R\backslash\{r\})-d(v,r))\right\}~.
\end{equation}
Note that
\begin{eqnarray}
\label{eq:CFT} \nonumber C_{RFTFL}(I,R) &=&
C_{facil}(I,R) + \max_{r\in R}
\left\{ C_{ship}(I, R\setminus \{r\})\right\} \\ \nonumber
&=& C_{facil}(I,R) +
\max_{r\in R} \left\{\sum_{v=1}^n{SC_{v,R\setminus\{r\}}} \right\}
\\
&=&
C_{facil}(I,R) + \max_{r\in R}
\left\{\sum_{v=1}^n \omega(v)\cdot d(v,R\backslash\{r\})\right\} ~.
\end{eqnarray}
Again, when the instance $I$ is clear from the context we omit it
and write simply $C_{RFTFL}(R)$, $C_{facil}(R)$, $C_{ship}(R)$,
$C_{backup}(R)$, etc.

We also consider the {\em robust $\alpha$-fault-tolerant facility
location ($\alpha\_$RFTFL)} problem, for integer $\alpha \geq 1$,
where the solution should be resilient against a failure of up to
$\alpha$ nodes. We define the $\alpha\_$RFTFL as follows. Each
client is supplied by the nearest open facility which did not
fail. We are looking for a subset $R \subseteq \{1,...,n\}$ that
minimizes the cost function
\begin{equation}
\label{eq:alpha_CFT} C_{\alpha\_RFTFL}(I,R) ~=~ C_{facil}(I,R) +
\max_{|R'| \leq \alpha} \left\{\sum_{v=1}^n \omega(v)\cdot
d(v,R\setminus R')\right\}~.
\end{equation}

\section{A constant approximation algorithm for RFTFL}

\subsection{The concentrated backup problem and its approximation}
\label{sec:cbp}

Towards developing a constant ratio approximation algorithm for
RFTFL, we first consider a different problem, named {\em
concentrated backup (conc\_bu)}, defined as follows. An instance
of the problem consists of a pair $\langle I,R_1\rangle$ where
$I=\langle G,l,f,\omega\rangle$ is defined as before and
$R_1=\{r_1,...,r_k\}$ is a set of nodes. In this version, the
nodes of $R_1$ act as both clients and servers (with open
facilities), and all other nodes $v \notin R_1$ have zero demands.
Informally, it is assumed that we have already paid for opening
the facilities in $R_1$, and each $r \in R_1$ serves itself, at
zero shipping cost. The problem requires to assign each client
$r\in R_1$ to a backup server $v \neq r$, which may be either some
server in $R_1$ or a new node from $V\backslash R_1$. For a set of
nodes $R_2$, define the {\em backup cost}
$$C_{bu}(I,R_1,R_2) =
\max\limits_{r \in R_1}\left\{SC_{r,R_1 \cup R_2 \setminus\{r\}}\right\} =
\max\limits_{r \in R_1}{\left\{\omega(r) d(r, R_1 \cup R_2\backslash
\{r\})\right\}}.$$
We are looking for a set $R_2$ minimizing
\begin{equation}
\label{eq:conc_bu}
C_{conc\_bu}(I,R_1,R_2)=C_{facil}(R_2)+C_{bu}(R_1,R_2).
\end{equation}
We denote this minimum cost by $C^*_{conc\_bu}(I,R_1)$.
We show a 2-approximation algorithm for the concentrated backup problem.

The problems studied in this section and in section \ref{sec:2.2.1}
are closely related to those considered in \cite{kps}, and to solve them
we use methods similar to the ones presented in \cite{kps}.

Let us consider a simpler variant of the backup problem, named
the {\em bounded backup (bb)} problem, which is defined on $\langle I, R_1, M
\rangle$ and requires looking for a solution $R_2$ minimizing
$$C_{bb}(I,R_1,M,R_2)= C_{facil}(R_2)$$
subject to the
constraint $C_{bu}(R_1,R_2) \leq M$, for integer $M$. We now
present a relaxation algorithm that finds a set $R_2$ satisfying
$C_{facil}(R_2) \leq C_{bb}^*(R_1,M)$ but obeying only the relaxed
constraint $C_{bu}(R_1,R_2) \leq 2M$ instead $C_{bu}(R_1,R_2) \leq M$.

\begin{figure}[htbp]
\begin{center}
\framebox{\hspace{0.0cm}\parbox{5.5in}{
{\boldmath Algorithm $A_{bb}(I, R_1, M)$}
\begin{enumerate}
\item
$R_{bb}^{alg} \gets \emptyset$
\item
For $i=1$ to $k$ do:
\begin{itemize}
\item
$S_i \gets \{v \mid \omega(r_i) d(v,r_i) \leq 2M\} \backslash \{r_i\}$
/* ``relaxed" backup servers for $r_i$ */
\item
If $S_i\cap (R_1\cup R_{bb}^{alg})=\emptyset$ then add to $R_{bb}^{alg}$
the node $v$ in $S_i$  with the minimum facility cost $f(v)$.
\end{itemize}
\item
Return $R_{bb}^{alg}$.
\end{enumerate}
}
\hspace*{0.6cm}}
\end{center}
\end{figure}

Let us now prove the properties of algorithm $A_{bb}$. For every
${r_i \in R_1}$ let the set of feasible backup servers be
$T_i=\{v\mid \omega(r_i)d(v,r_i)\leq M \} \backslash \{ r_i\}$.
Let the set of relaxed backup servers selected by the algorithm
(namely, the final set $R_{bb}^{alg}$ it returns) be
$R_{bb}^{alg}(R_1,M) = \{q_1^{alg},...,q_J^{alg}\}$. Let $\ell_j$
be the phase in which the algorithm adds the new
facility $q_j^{alg}$ to $R_{bb}^{alg}$, for $1 \leq j \leq J$.

\begin{lemma}
\label{lem:3.1} $T_{\ell_i} \cap T_{\ell_j} = \emptyset$ for $1
\leq i,j \leq J$.
\end{lemma}
\Proof
Assume otherwise, and let $v \in T_{\ell_i} \cap
T_{\ell_j}$ for some $1 \leq i,j \leq J, i \neq j$. Assume without
loss of generality that $\omega(r_{\ell_i}) \leq
\omega(r_{\ell_j})$. Since $\omega(r_{\ell_j})d(v,r_{\ell_j}) \leq
M$, necessarily $\omega(r_{\ell_i})d(v,r_{\ell_j}) \leq M$ as
well, and by the definition of $T_{\ell_i}$, also
$\omega(r_{\ell_i})d(v,r_{\ell_i}) \leq M$, hence
$$\omega(r_{\ell_i})d(r_{\ell_i},r_{\ell_j}) ~\leq~
\omega(r_{\ell_i})(d(v,r_{\ell_i}) + d(v,r_{\ell_j})) ~\leq~ 2M,$$
implying that $r_{\ell_j} \in S_{\ell_i} \cap R_1$, so the
algorithm should not have opened a new facility in phase
$\ell_i$, contradiction. \QED

\begin{lemma}
\label{lem:3.3} $C_{facil}(R_{bb}^{alg}(R_1,M)) \leq C_{bb}^*(
R_1,M)$.
\end{lemma}
\Proof
Notice that there must be at least one node from every
$T_i$ in the optimal solution $R_{bb}^*(R_1,M)$. By Lemma
\ref{lem:3.1} the sets $T_{\ell_1},...,T_{\ell_J}$ are disjoint, so
there are at least $J$ distinct nodes $q^*_{j} \in R_{bb}^*(R_1,M)$,
one from each
$T_{\ell_j}$, for $1 \leq j \leq J$. In each phase $i$,
the algorithm selects the
cheapest node in $S_i \supseteq T_i$. Therefore,
$f(q_j^{alg}) \leq f(q_{j}^*)$ for every $1 \leq j \leq J$.
Hence $C_{facil}(R_{bb}^{alg}(R_1,M)) =
\sum\limits_{j=1}^J{f(q_j^{alg})} \leq
\sum\limits_{j=1}^J{f(q_{j}^*)} \leq C_{bb}^*(R_1,M).$ \QED

\begin{lemma}
\label{lem:3.4} $C_{bu}(R_1,R_{bb}^{alg}(R_1,M)) \leq 2M$.
\end{lemma}
\Proof
For each server $r_i$ in $R_1$, the algorithm ensures that there
is at least one open facility from the set $S_i$, so
$\omega(r_i)d(r_i,R_1 \cup R_{bb}^{alg}(R_1,M) \setminus \{r_i\}) \leq 2M$.
\QED

Now we present an approximation algorithm $A_{conc\_bu}$
for the concentrated backup problem using the
relaxation algorithm $A_{bb}$ for the bounded backup problem.
First note that there can be at most $nk$ possible values
for the shipping costs $SC_{u,v}=\omega(u) d(u,v)$.

\begin{figure}[htbp]
\begin{center}
\framebox{\hspace{0.0cm}\parbox{5.5in}{
{\boldmath Algorithm $A_{conc\_bu}(I, R_1)$}
\begin{enumerate}
\item
For every $M \in \{SC_{u,v}\mid u,v \in V\}$ do:
\begin{itemize}
\item
let $R_{bb}^{alg}(R_1,M) \gets A_{bb}(I, R_1, M)$.
\end{itemize}
\item
Return the set $R_{bb}^{alg}(R_1,M)$ with the minimum cost
$C_{conc\_bu}(R_1,R_{bb}^{alg}(R_1,M))$.
\end{enumerate}
}
\hspace*{0.6cm}}
\end{center}
\end{figure}

\begin{lemma}
\label{lem:3.5} $C_{conc\_bu}^{alg}(I,R_1) \leq
2C_{conc\_bu}^*(I,R_1).$
\end{lemma}
\Proof
Recall that, letting $R_2^*=R_{conc\_bu}^*(R_1)$,
\begin{eqnarray*}
C_{conc\_bu}^*(I,R_1) &=& C_{conc\_bu}(I,R_1,R_2^*) ~=~
C_{facil}(R_2^*)+ C_{bu}(I,R_1,R_2^*).
\end{eqnarray*}
Let $u \in R_1$ be the node that attains the maximum shipping cost
$SC_{u,R_1\cup R_2 \setminus \{u\}}$, i.e., satisfies $\omega(u)
d(u,R_1\cup R_2^* \backslash \{u\})=C_{bu}(I, R_1,R_2^*)$, and let $v
\in R_1\cup R_2^* \backslash \{u\}$ be its backup,
i.e., the closest node to $u$.
Then $C_{conc\_bu}^*(I,R_1)=C_{conc\_bu}(I,R_1,R_2^*)=
C_{facil}(R_2^*)+ SC_{u,v}$. Since the algorithm examines all
possible values of $M$, it tests also $M_0 = SC_{u,v}$. For this
value, the returned set $R_{bb}^{alg}(R_1,M_0)$ has opening cost
at most $C_{bb}^*(R_1,M_0) = C_{facil}(R_2^*)$ and backup cost at most
\\
$C_{bu}(I,R_1,R_{bb}^{alg}(R_1,M_0)) \leq 2M_0$ by Lemmas
\ref{lem:3.3} and \ref{lem:3.4}. Since the algorithm takes the minimum cost
$C_{conc\_bu}(R_1,R_{bb}^{alg}(R_1,M))$ over all possible values of $M$,
the resulting cost satisfies $C_{conc\_bu}^{alg}(I,R_1)
\leq C_{facil}(R_2^*) + 2SC_{u,v} \leq 2C_{conc\_bu}^*(I,R_1)$,
namely, an approximation ratio of 2.
\QED

\subsection{6.5-approximation algorithm for RFTFL}

We now present a polynomial time algorithm $A_{RFTFL}$ that yields
6.5-approximation for the robust fault-tolerant uncapacitated facility location
problem RFTFL. Consider an
instance $I=\langle G ,l ,f, \omega\rangle$ of the problem. The
algorithm consists of three stages.

\noindent{\bf Stage 1:}
Apply the 1.5-approximation algorithm of \cite{jb} to the original UFL
problem in order to find an initial subset $R_1$ of servers. Notice that
the cost of this solution satisfies
\begin{equation}
\label{eq:UFL_RFTFL} C_{UFL}(R_1) \leq 1.5C_{UFL}^*\leq
1.5C_{RFTFL}^{*}~.
\end{equation}
Each node is now assigned to a server in $R_1$.
Next, we need to assign to each node a {\em backup server} which
will serve it in case its original server fails.

\noindent{\bf Stage 2:} Transform the given instance $I=\langle V,
l, \omega,f\rangle$ of the problem into an instance $I'=\langle
V,l,\omega',f'\rangle$ as follows. First, change the
facility cost $f$  by setting $f'(r)=0$ for $r \in R_1$. Next,
for each server $r \in R_1$, relocate all the demands of the nodes
that are served by $r$, and place them at the server $r$ itself,
that is, set
\begin{equation}
\omega'(r) = \left\{
\begin{array}{rl}
\sum\limits_{v \in \varphi(I, r,R_1)}{\omega(v)}, & ~for~ r \in R_1,\\
0, & ~for~ r \notin R_1.
\end{array} \right.
\end{equation}

\noindent{\bf Stage 3:} Invoke the 2-approximation
algorithm $A_{conc\_bu}$ for the concentrated backup problem on
the new instance $I'$ and the set $R_1$. The approximation
algorithm returns a new set $R_2$. We then return the set $R_1
\cup R_2$ as the final set of open facilities.

\begin{lemma}
\label{lem:3.6} For every instance $I$ and set $R_1
\subseteq V$, $C_{conc\_bu}^*(I',R_1)\leq C_{RFTFL}^*(I) +
C_{UFL}(I,R_1)$.
\end{lemma}
\Proof
Consider some vertex $r \in R_1$ and let
$\varphi(I,r,R_1)=\{v_1^r,...,v_{k_r}^r\}$ be the nodes it serves.
Consider the optimal solution $R_{RFTFL}^*(I)$ to the RFTFL problem.
Let $d_i^r$ be the distance from $r$ to $v_i^r$ for $1 \leq i \leq
k_r$, and also let $x_i^r$ be the distance from $v_i^r$ to its
optimal backup server, which is also its distance to $R_r^* \equiv R_1 \cup
R_{RFTFL}^*(I) \backslash \{r\}$, i.e., $x_i^r=d(v_i^r, R_r^*)$.
By the triangle inequality, $d(r,R_r^*) \leq d_i^r + x_i^r$, for every
$1 \leq i \leq k_r,$ so
\begin{eqnarray*}
\omega'(r) \cdot d(r,R_r^*) &=&
\sum_{l=1}^{k_r}{\omega(v_l^r)} \cdot d(r,R_r^*) \leq
\sum_{l=1}^{k_r}{\omega(v_l^r)(d_l^r + x_l^r)} \\ &=&
\sum_{l=1}^{k_r}{\omega(v_l^r)} d(v_l^r, R_1) +
\sum_{l=1}^{k_r}{\omega(v_l^r) x_l^r} \\ &\leq&
\sum_{v=1}^{n}{\omega(v) \cdot d(v,R_1)} +
\sum_{v=1}^{n}{\omega(v) \cdot d(v, R_r^*)}.
\end{eqnarray*}
Therefore,
\begin{eqnarray*}
C_{bu}(I',R_1,R^*_{RFTFL}(I)) &=& \max\limits_{r\in R_1}
{\left\{\omega'(r) \cdot d(r,R_r^*)\right\}} \\
&\leq& C_{ship}(I,R_1) + \max\limits_{r \in R_1}
{\left\{\sum\limits_{v=1}^{n}{\omega(v) \cdot d(v, R_r^*)}\right\}}.
\end{eqnarray*}
Using (\ref{eq:CFT}) and (\ref{eq:conc_bu}) we now bound the cost
of the optimal solution for problem $conc\_bu$ by
\begin{eqnarray*}
C_{conc\_bu}^*(I',R_1)
&\leq& C_{conc\_bu}(I',R_1, R_{RFTFL}^*(I))
\\ &=& C_{facil}(I',R_{RFTFL}^*(I))+ C_{bu}(I',R_1,R^*_{RFTFL}(I))
\\
&\leq& C_{facil}(I',R_{RFTFL}^*(I)) + \max\limits_{r \in
R_1}{\left\{\sum\limits_{v=1}^n{\omega(v) d(v,R_r^*) }\right\}}
~+~ C_{ship}(I,R_1)
 \\
&\leq& C_{RFTFL}^*(I)+C_{ship}(I,R_1)
~\leq~ C_{RFTFL}^*(I,R_1)+C_{UFL}(I,R_1). \inQED
\end{eqnarray*}

\begin{lemma}
\label{lem:2.7}
For every instance $I$ and sets $R_1, R_2 \subseteq V$,

$C_{RFTFL}(I,R_1 \cup R_2) \leq C_{UFL}(I,R_1) + C_{conc\_bu}(I',R_1,R_2)$.
\end{lemma}
\Proof
The cost of opening the facilities in $R_1 \cup R_2$ is clearly at most
the cost of opening the facilities in $R_1$ plus the cost of opening
the facilities in $R_2$.
For every facility $r \in R_1 \cup R_2$, in order to bound
$C_{ship}(I, R_1 \cup R_2 \setminus \{r\})$, note that one can first
move each client $v$ to its closest open facility in $R_1$,
and then move all the clients assigned to $r$ (if $r \in R_1$)
to the backup facility of $r$ in $R_2$. The inequality follows.
More formally we have the following.
Recall that by (\ref{eq:CFT}),
$$C_{RFTFL}(I,R_1 \cup R_2) = C_{facil}(I,R_1 \cup R_2)
+ \max_{r\in R_1 \cup R_2}\left\{\
C_{ship}(I,R_1 \cup R_2 \setminus \{r\})\right\}.$$
Consider first the case that
$\max\limits_{r\in R_1 \cup R_2}\left\{C_{ship}(I,R_1 \cup R_2
\setminus \{r\}) \right\}$ is attained for some $r' \in R_2$.
In this case, we get by (\ref{eq:1}) that
\begin{eqnarray*}
C_{RFTFL}(I,R_1 \cup R_2)
&=& C_{facil}(I,R_1 \cup R_2)
+ C_{ship}(I, R_1 \cup R_2 \setminus \{r'\})
\\
&\leq&
C_{facil}(I,R_1 \cup R_2) + C_{ship}(I,R_1)
\\
&=&
C_{UFL}(I,R_1) + C_{facil}(I,R_2) \leq  C_{UFL}(I,R_1)
+ C_{conc\_bu}(I',R_1,R_2).
\end{eqnarray*}
So now assume that $\max\limits_{r\in R_1 \cup R_2}
\left\{C_{ship}(I,R_1\cup R_2 \setminus \{r\}) \right\}$
is attained for some $r' \in R_1$. Therefore,
\begin{eqnarray*}
C_{RFTFL}(I,R_1 \cup R_2)
 &=& C_{facil}(I,R_1 \cup R_2) + C_{ship}(I,R_1 \cup R_2 \setminus \{r'\})
\\
&=&
C_{facil}(I,R_1) + C_{facil}(I,R_2) + \sum_{v=1}^n
{SC_{v,R_1 \cup R_2}}
\\
&&+ \sum_{v\in\varphi(I,r',R_1 \cup R_2)}
\omega(v)\cdot(d(v,R_1 \cup R_2\backslash\{r'\})-d(v,r'))
\\ &\leq&
C_{UFL}(I,R_1) + C_{facil}(I,R_2) \\ &&
+ \max_{r\in R_1}\left\{\sum_{v\in\varphi(I,r,R_1)}
\omega(v)\cdot(d(r,R_1 \cup R_2\backslash\{r\}))\right\}
\\ &=&
C_{UFL}(I,R_1) + C_{facil}(I,R_2)
+ \max_{r\in R_1}\left\{w'(r)\cdot(d(r,R_1 \cup R_2\backslash\{r\}))\right\}
\\ &=&
C_{UFL}(I,R_1) + C_{conc\_bu}(I',R_1,R_2).
\inQED
\end{eqnarray*}

\begin{lemma}
Algorithm $A_{RFTFL}$ yields a 6.5-approximation for the RFTFL
problem.
\end{lemma}
\Proof
Consider the set of opened facilities $R_1 \cup R_2$.
By Lemma \ref{lem:3.5}, $R_2$ is a 2-approximation of the concentrated
backup problem on the instance $I'$, so
$$C_{conc\_bu}(I',R_1,R_2) \leq 2C_{conc\_bu}^*(I',R_1).$$
By Lemma \ref{lem:3.6},
$C_{conc\_bu}^*(I',R_1) \leq C_{RFTFL}^*(I) + C_{UFL}(I,R_1)$,
hence
$$C_{conc\_bu}(I',R_1,R_2) \leq 2 C^*_{RFTFL}(I) +
2 C_{UFL}(I,R_1).$$
Using Lemma \ref{lem:2.7} we get
$$C_{RFTFL}(I,R_1 \cup R_2) ~\leq~
3C_{UFL}(I,R_1) + 2C_{RFTFL}^*(I),$$
and by (\ref{eq:UFL_RFTFL}),
$C_{RFTFL}(I,R_1 \cup R_2) ~\leq~ 6.5C_{RFTFL}^*(I)$. \QED

\section{An approximation algorithm for $\alpha\_$RFTFL}

\subsection{The concentrated $\alpha\_$backup problem}
\label{sec:2.2.1}

As in the case of a single failure, we first consider a different problem,
named
{\em concentrated $\alpha\_$backup ($conc\_\alpha\_bu$)}, defined as follows.
An instance of the problem consists of a pair $\langle
I,R_1\rangle$ where $I=\langle G,l,f,\omega\rangle$ is defined as
before and $R_1$ is a set of nodes. The nodes of $R_1$ act as both
clients and servers (with open facilities), and all other nodes $v
\notin R_1$ have zero demands. We are looking for a set $R_2$
minimizing
\begin{equation}
\label{eq:conc-alpha-bu} C_{conc\_\alpha\_bu}(I,R_1,R_2) =
C_{facil}(R_2)+ C_{\alpha\_bu}(I,R_1,R_2),
\end{equation}
where $C_{\alpha\_bu}$ is the {\em maximum $\alpha\_$backup cost} for a set
of nodes $R_2$, defined as
$$C_{\alpha\_bu}(I,R_1,R_2) = \max\limits_{|F| \leq
\alpha}{\left\{\sum\limits_{r \in (F \cap R_1)}{\omega(r) \cdot d(r,R_1 \cup
R_2 \backslash F)}\right\}}.$$

We will shortly present a $3\alpha$-approximation algorithm for the
concentrated $\alpha$-backup problem.

Towards this, let us first consider a simpler variant of the backup problem,
named
the {\em $\alpha$-bounded backup ($\alpha\_bb$)} problem, which is defined on
$\langle I, R_1, M\rangle$ and requires looking for a solution $R_2$
minimizing
$$C_{\alpha\_bb}(R_1,M,R_2)= C_{facil}(R_2)$$
subject to the constraint $C_{light\_\alpha\_bu}(R_1,R_2) \leq M$
for some integer $M$, where
$$C_{light\_\alpha\_bu}(R_1,R_2) = \max\limits_{r \in R_1, |F| \leq
\alpha }{\left\{\omega(r) d(r,R_1 \cup R_2 \backslash F)\right\}}.$$

We now
present a relaxation algorithm that finds a set $R_2$ satisfying
$C_{facil}(R_2) \leq C_{\alpha\_bb}^*(R_1,M)$ but allowing the
relaxed constraint $C_{light\_\alpha\_bu}(R_1,R_2) \leq 3M$ instead of
$C_{light\_\alpha\_bu}(R_1,R_2) \leq M$.

\begin{figure}[htbp]
\begin{center}
\framebox{\hspace{0.0cm}\parbox{5.5in}{
{\boldmath Algorithm $A_{\alpha\_bb}(I, R_1,M)$}
\begin{enumerate}
\item
$R_{\alpha\_bb}^{alg} \gets \emptyset$
\item
Let ${r_1,...,r_k}$ be the servers in $R_1$ sorted by nonincreasing order
of demands.
\item
$Z \gets  \emptyset$
/* The set of servers $r_i$ where the algorithm opens facilities
in phase $i$ */
\item
For $i=1$ to $k$ do:
\item
\begin{itemize}
\item
$S_i \gets \{v \mid \omega(r_i) d(v,r_i) \leq 2M\} \backslash
\{r_i\}$.
\item
$T_i \gets \{v\mid \omega(r_i)d(v,r_i)\leq M \}
\backslash \{r_i\}$
\item
If $S_i \cap Z  = \emptyset$ then:
\begin{itemize}
\item
Add to $R_{\alpha\_bb}^{alg}$, the $\alpha - |T_i \cap (R_1 \cup
R_{\alpha\_bb}^{alg})|$ nodes in $T_i \backslash (R_1 \cup
R_{\alpha\_bb}^{alg})$ with the lowest facility costs.
\item
$Z \gets Z \cup \{r_i\}$
\end{itemize}
\end{itemize}
\item
Return $R_{\alpha\_bb}^{alg}$.
\end{enumerate}
}
\hspace*{0.6cm}}
\end{center}
\end{figure}

Let us now prove the properties of Alg. $A_{\alpha\_bb}$. Let $\{
\ell_j \mid 1 \leq j \leq J \}$ be the phases in which the
algorithm adds new facilities to $R_{\alpha\_bb}^{alg}$.
By a proof similar to that of Lemma \ref{lem:3.1}, we have the following.

\begin{lemma}
\label{lem:4.1} $T_{\ell_i} \cap T_{\ell_j} = \emptyset$ for $1
\leq j < i \leq J$.
\end{lemma}

\begin{lemma}
\label{lem:4.3} $C_{facil}(R_{\alpha\_bb}^{alg}(R_1,M)) \leq
C_{\alpha\_bb}^*(R_1,M)$.
\end{lemma}
\Proof
There must be at least $\alpha$ nodes in every $T_{\ell_j}$ in the optimal
solution $R_{\alpha\_bb}^*(R_1,M)$. By Lemma \ref{lem:4.1} the
sets $T_{\ell_j}$ for $1 \leq j \leq J$ are disjoint, so the only
nodes that the algorithm adds to $R_{\alpha\_bb}^{alg}$ from the set
$T_{\ell_j}$ are added at phase
$\ell_j$. The algorithm selects the cheapest nodes in
$T_{\ell_j}$ in order to complete to $\alpha$ nodes. Therefore,
$C_{facil}(R_{\alpha\_bb}^{alg}(R_1,M) \cap T_{\ell_j})
\leq C_{facil}(R_{\alpha\_bb}^*(R_1,M) \cap T_{\ell_j})$
for every $1 \leq j \leq J$. Hence
\begin{eqnarray*}
C_{facil}(R_{\alpha\_bb}^{alg}(R_1,M)) &=&
\sum\limits_{j=1}^J{C_{facil}(R_{\alpha\_bb}^{alg}(R_1,M) \cap
T_{\ell_j})} ~\leq~
\sum\limits_{j=1}^J{C_{facil}(R_{\alpha\_bb}^{*}(R_1,M) \cap
T_{\ell_j})} \\ &\leq& C_{\alpha\_bb}^*(R_1,M). \inQED
\end{eqnarray*}

\begin{lemma}
\label{lem:4.4} $C_{\alpha\_bu}(R_1,R_{\alpha\_bb}^{alg}(R_1,M))
\leq 3M$.
\end{lemma}
\Proof
For each server $v_i \in R_1$, the algorithm ensures that
either there are at least $\alpha$ open facilities from the set $T_i$ or
$v_i$ is at distance at most $2M$ from another $v_j \in R_1$ that
has $\alpha$ open facilities from the set $T_j$. In the first case the
distance is at most $M$ and in the second - at
most $3M$. \QED

Now we present an approximation algorithm $A_{conc\_\alpha\_bu}$ for the
concentrated $\alpha\_$backup problem, using the relaxation algorithm
$A_{\alpha\_bb}$ for the $\alpha\_$bounded backup problem.

\begin{figure}[htbp]
\begin{center}
\framebox{\hspace{0.0cm}\parbox{5.5in}{
{\boldmath Algorithm $A_{conc\_\alpha\_bu}(I, R_1)$}
\begin{enumerate}
\item
For every subset $T \subseteq \{SC_{v,u}\mid v,u \in V \}$ such that
$|T| \leq \alpha$ do:
\begin{itemize}
\item
$M(T) \gets \sum_{m \in T}{m}$
\item
let $R_{\alpha\_bb}^{alg}(R_1,M(T)) \gets A_{\alpha\_bb}(I,R_1, M(T))$.
\end{itemize}
\item
Return the set $R_{\alpha\_bb}^{alg}(R_1,M(T))$ with the minimum cost \\
$C_{conc\_\alpha\_bu}(R_1,R_{\alpha\_bb}^{alg}(R_1,M(T)))$.
\end{enumerate}
}
\hspace*{0.6cm}}
\end{center}
\end{figure}

\begin{lemma}
\label{lem:4.5} $C_{conc\_\alpha\_bu}^{alg}(I, R_1) \leq 3\alpha
C_{conc\_\alpha\_bu}^*(I, R_1).$
\end{lemma}
\Proof
Denote the optimal solution for $conc\_\alpha\_bu$ on $\langle I, R_1 \rangle$
by $R_2^*=R_{conc\_\alpha\_bu}^{*}(R_1)$. Then
\begin{eqnarray*}
C_{conc\_\alpha\_bu}^*(I,R_1) = C_{conc\_\alpha\_bu}(I,R_1,R_2^*)
= C_{facil}(R_2^*)+
C_{\alpha\_bu}(I, R_1,R_2^*).
\end{eqnarray*}
Let $\{u_1,...,u_j\} \subseteq R_1$ and $\{v_1,...,v_j\} \subseteq
R_1\cup R_2^*$ for some $j \leq \alpha$ be the sets of nodes
that attain the maximum shipping cost, i.e.,
satisfy $C_{\alpha\_bu}(I, R_1,R_2^*) = M_0$
for $M_0 = \sum\limits_{i=1}^j{SC_{u_i, v_i}}=
\sum\limits_{i=1}^j{\omega(u_i) d(u_i, v_i)}$.  Then
$C_{conc\_\alpha\_bu}^*(I,R_1) = C_{facil}(R_2^*) + M_0$.
Notice that there must be at least $\alpha$ nodes
in the set $R_2^* \cup R_1$ at distance at most $M_0$ from every server $r$
in $R_1$. Clearly
$C_{facil}(R_{\alpha\_bb}^*(R_1,M_0)) \leq C_{facil}(R_2^*)$.
Since the algorithm examines all possible values of $M(T)$, it
tests also $M_0$. For this value, the returned set
$R_{\alpha\_bb}^{alg}(R_1,M_0)$ has opening cost at most
$C_{\alpha\_bb}^*(R_1,M_0) \leq C_{facil}(R_2^*)$ and
backup cost at most
$C_{\alpha\_bu}(I,R_1,R_{\alpha\_bb}^{alg}(R_1,M_0)) \leq 3M_0$ by
Lemmas \ref{lem:4.3} and \ref{lem:4.4}. Since the algorithm takes the minimum
cost $C_{conc\_\alpha\_bu}(R_1,R_{\alpha\_bb}^{alg}(I,R_1,M(T)))$
over all possible subsets $T$, the resulting cost is at most
\begin{eqnarray*}
C_{conc\_\alpha\_bu}^{alg}(I,R_1) &\leq&
C_{conc\_\alpha\_bu}(I,R_1,
R_{\alpha\_bb}^{alg}(R_1,M_0)) \\
&\leq& C_{facil}(R_2^*) + \max\limits_{|F| \leq
\alpha}{\left\{\sum\limits_{r \in (F \cap R_1)}{\omega(r) d(r,R_1 \cup
R_{\alpha\_bb}^{alg}(R_1,M_0) \backslash F)}\right\}} \\
&\leq&
C_{facil}(R_2^*)+ 3 \alpha M_0 ~\leq~
 3\alpha C_{conc\_\alpha\_bu}^*(I,R_1)
.~\inQED
\end{eqnarray*}

\subsection{$(1.5 + 7.5\alpha)$-approximation algorithm to the $\alpha\_$RFTFL}

We now present a polynomial time algorithm named $A_{\alpha\_RFTFL}$, yielding
a $(1.5 + 7.5\alpha)$-approximation for the robust fault-tolerant uncapacitated
facility location problem $\alpha\_$RFTFL against a failure of $\alpha$ nodes,
for constant $\alpha > 1$.
Consider an instance $I=\langle G ,l ,f, \omega\rangle$ of the problem.
The algorithm is similar to Algorithm RFTFL, except for the third stage.
Instead of invoking the 2-approximation algorithm $A_{conc\_bu}$
for the concentrated backup problem on the new instance $I'$ and
the set $R_1$, invoke the $3\alpha$-approximation algorithm
$A_{conc\_\alpha\_bu}$ for the concentrated $\alpha\_$backup problem
on the new instance $I'$ and the set $R_1$. Algorithm $A_{conc\_\alpha\_bu}$
returns a new set $R_2^{alg}$.
Algorithm $A_{\alpha\_RFTFL}$ now returns the set $R_1 \cup R_2^{alg}$.
Proof of the following lemma is deferred to the full paper.

\begin{lemma}
\label{lem:4.7}
Algorithm $A_{\alpha\_RFTFL}$ yields a $(1.5 + 7.5\alpha)$-approximation for 
the $\alpha\_$RFTFL problem.
\end{lemma}

\section{Robust Fault-tolerant uncapacitated facility location on trees}

In this section we show that the RFTFL problem is NP-hard even on trees.
The claim holds even in the case where only the edge lengths or only
the node demands
are variable and the other parameters are uniform. An instance of the
RFTFL problem is $\langle T,l,f,\omega,P \rangle$, where $T$ is a
tree, $l,f$ and $\omega$ are defined as before and $P$ is an integer.
It is required to decide if the cost of the optimal
solution to the RFTFL problem on the instance $\langle T,l,f,\omega
\rangle$ is $P$ or less.

The proofs, via reductions from
subset sum and from a variant of the partition problem,
are deferred to the full paper.
The following results are established.

\begin{theorem}
\label{thm:NPhard}
$RFTFL$ on trees is NP-complete even with
\begin{enumerate}
\item
unit edge lengths and opening costs (but variable node demands),
\item
unit node demands and opening costs (but variable edge lengths).
\end{enumerate}
\end{theorem}


\end{document}